# The Reclassification of Asteroids from Planets to Non-Planets

Philip T. Metzger, Florida Space Institute, University of Central Florida, 12354 Research Parkway, Partnership 1 Building, Suite 214, Orlando, FL 32826-0650; philip.metzger@ucf.edu

Mark V. Sykes, Planetary Science Institute, 1700 E. Fort Lowell, Suite 106, Tucson, AZ 85719; sykes@psi.edu.

Alan Stern, Southwest Research Institute, 1050 Walnut St, Suite 300, Boulder, CO 80302; alan@boulder.swri.edu

Kirby Runyon, Johns Hopkins University Applied Physics Laboratory, Laurel, MD 20723, USA.; kirby.runyon@gmail.com

## Abstract

It is often claimed that asteroids' sharing of orbits is the reason they were re-classified from planets to non-planets. A critical review of the literature from the 19[th] Century to the present shows this is factually incorrect. The literature shows the term *asteroid* was broadly recognized as a subset of *planet* for 150 years. On-going discovery of asteroids resulted in a *de facto* stretching of the concept of *planet* to include the ever-smaller bodies. Scientists found utility in this taxonomic identification as it provided categories needed to argue for the leading hypothesis of planet formation, Laplace's nebular hypothesis. In the 1950s, developments in planet formation theory found it no longer useful to maintain taxonomic identification between asteroids and planets, Ceres being the primary exception. At approximately the same time, there was a flood of publications on the geophysical nature of asteroids showing them to be geophysically different than the large planets. This is when the terminology in asteroid publications calling them *planets* abruptly plunged from a high level of usage where it had hovered during the period 1801 – 1957 to a low level that held constant thereafter. This marks the point where the community effectively formed consensus that asteroids should be taxonomically distinct from planets. The evidence demonstrates this consensus formed on the basis of geophysical differences between asteroids and planets, not the sharing of orbits. We suggest attempts to build consensus around planetary taxonomy not rely on the non-scientific process of voting, but rather through precedent set in scientific literature and discourse, by which perspectives evolve with additional observations and information, just as they did in the case of asteroids.

## Introduction

The claim that asteroids were reclassified from the planet taxon into a non-planet category because of orbit sharing or existing in large numbers has been widely and uncritically repeated in popular media since 2006 (e.g., Overbye, 2006); it begs the question whether this really was the historical reason why asteroids were reclassified as non-planets, and yet no historical evidence is ever provided. Could it not have been another, possibly geophysical, reason? Is this widespread reporting merely urban legend based on a bad understanding of the scientific process used in the 19[th] and 20[th] centuries?

The website of the International Astronomical Union (IAU) says,

> In the 19th century astronomers could not resolve the size and shape of Ceres, **and because numerous other bodies were discovered in the same region**, Ceres lost its planetary status. (IAU, 2018, bold added)

Soter (2006) wrote,

> Between 1845 and 1851, the population of known asteroids increased to 15, and the continued planetary status of these small bodies became unwieldy. Astronomers then began to number all asteroids by their order of discovery, rather than by semimajor axis, as for planets…This marked the de facto acceptance of the asteroids as members of a population distinct from planets.

But are these claims true? Did astronomers really take away Ceres' planetary status "because numerous other bodies were discovered"? Did astronomers really share "acceptance" that asteroids are no longer part of the *planet* taxon because there were as many as 15? Numeration is not a taxonomic definition. A thorough review of the 19[th], 20[th] and 21[st] century literature reveals a clear and interesting story of the extent to which asteroids were considered a class of planet over 150 years that is at variance with the IAU position.

## The Only Case

We found that William Herschel's 1802 paper (Herschel, 1802) is the only case in the literature where a scientist made an argument about non-sharing of orbits to reclassify asteroids, prior to the recent dispute leading to the IAU vote in 2006. After the first two asteroids Ceres and Pallas were discovered, Herschel presented a list of geophysical and dynamical properties that suggest these bodies are a fundamentally different type of body than the previously known planets and comets, so he proposed calling them asteroids, meaning "star-like objects." (This was because their tiny size made it impossible to resolve their disks, just as with stars that are too far away.) Herschel pointed out the inclination of these bodies is much higher than the inclinations of the other known planets at that time, and if they are planets then "we should be obliged to give up the zodiac." Herschel argued planets may have moons but asteroids do not, and he argued asteroids have atmospheric comas much larger than any planetary atmospheres (which was just an observational error probably due to his equipment). Herschel's last of five arguments against their *planet* classification was that their orbital spacing was not as regular as that of the known planets, and he claimed the elegance of the Titius-Bode Law (Titius, 1766) would be destroyed if we admitted them as planets:

> … for it appears, that their orbits are too near each other to agree with the general harmony that takes place among the rest; perhaps one of them might be brought in, to fill up a seeming vacancy between Mars and Jupiter. There is a certain regularity in the arrangement of planetary orbits, which has been pointed out by a very intelligent astronomer, so long ago as the year 1772 [see Bode, 1772]; but this, by the admission of the two new stars into the order of planets, would be completely overturned; whereas, if they are of a different species, it may still remain established. (Herschel, 1802)

Most of W. Herschel's arguments are no longer considered relevant. He measured the sizes of Ceres and Pallas by factors of 2 to 3.7 smaller in diameter than they actually are. For Ceres, this difference is vital for it to be large enough for gravitational rounding, and most now consider it not just an asteroid but also a dwarf planet, or simply, planet (e.g., Hantzsche, 1996; McCord and Sotin, 2003; McCord, 2013). In other words, for Ceres the size is not a debarring property. Inclination is no longer considered relevant since it is recognized that planets can scatter one another and yet remain planets. For example, the so-called "Planet Nine" is believed to have a high inclination and high eccentricity (Brown and Batygin, 2016) yet would not be debarred from the planet taxon on that basis. Herschel's argument about moons was incorrect, because we now know there are very many asteroids with moons (e.g., Merline et al., 2002) and KBOs with moons (e.g., Christy and Harrington, 1980; Brown et al., 2006). Herschel's argument about comas of asteroids was based on an observational error, but even if it were not an error the existence of a large atmosphere is not considered a debarring property from the planet taxon. The last of Herschel's arguments was a version of the one in contention between the IAU's 2006 definition (IAU, 2006), which requires dynamical orbit clearing, in contrast with a geophysical definition that is based solely on the intrinsic properties of a body (Sykes, 2007; Sykes 2008, Runyon, et al. 2017). However, basing an orbit-clearing argument on the elegance of the Titius-Bode Law is no longer considered valid.

In his three subsequent discussions of the asteroids (Herschel, 1803; 1805; 1807), W. Herschel abandoned the spurious argument about them having comas, and he progressively retreated from the Titius-Bode argument. In 1803 he mentioned only Ceres' and Pallas' size and inclination,

> It is not in the least material whether we call them asteroids, as I have proposed; or planetoids, as an eminent astronomer, in a letter to me, suggested; or whether we admit them at once into the class of our old seven large planets. In the latter case, however, we must recollect, that if we would speak with precision they should be called **very small**, and **exzodiacal**; for, the great inclination of the orbit of one of them to the ecliptic, amounting to 35 degrees, is certainly remarkable. That of the other is also considerable; its latitude, the last time I saw it, being more than 15 degrees north. These circumstances, added to their smallness, show that there exists a greater variety of arrangement and size among the bodies which our sun holds in subordination… (Herschel, 1803, bold added)

In 1804 when Harding discovered Juno, Herschel mentioned four properties, "by its similar situation between Mars and Jupiter, as well as by the smallness of its disk, added to the considerable inclination and excentricity of its orbit, departs from the general condition of planets," but then he listed only two properties—size and inclination—as the ones that would have caused him to reconsider their status as planets: "by some approximation of a motion in the zodiac, or by a magnitude not very different from a planetary one" (Herschel, 1805). Upon Olbers' discovery of Vesta in 1807, Herschel mentioned again only size and inclination as the distinguishing features, "…their smallness and considerable deviation from the path in which the planets move…" (Herschel, 1807).

See Cunningham (2016) for a review of the contemporaneous response to W. Herschel's arguments. Banks, Huth, Laplace, Olbers, Piazzi, Regnér, and Zach believed various of Herschel's criteria should not carry weight in defining the *planet* taxon (Cunningham, 2016). Baron Franz von Zach generalized the question of which criteria are important:

> It will depend on what astronomers regard as the essential characteristics of planets: a path determined to be near-circular? Or the volume? (Zach, 1802)

*The Critical Review* did not believe Herschel adequately justified a new taxonomical system based on his expanded set of characteristics:

> Why then are not these bodies planets? We see no reason for any distinction: they revolve round the sun, and are not comets. We must discover another system, before we are allowed to change the appellation. (*The Critical Review,* 1803)

*The Annual Review, or Register of Literature* proposed that Herschel's additional characteristics are not essential to the *planet* taxon but define subcategories of the taxon:

> …it has been suggested…that there cannot be any need of a new name, and that the only distinction necessary is to divide the planets into two classes, zodiacal and extra-zodiacal. (*The Annual Review*, 1803)

In a future paper we will address how taxonomy relates to the scientific process and therefore why these taxonomical judgements matter. Here we focus on the narrower, preliminary question: did scientists really favor orbit clearing as an essential characteristic for planet taxonomy and thus reclassify asteroids as non-planets on that basis? We have found no support for this idea. The following literature review discovers the following seven trends that *do* exist in the two centuries from Herschel until now.

## 1. <u>Asteroids Were Planets</u>

The published literature shows that even after dozens, then hundreds and thousands of asteroids were discovered, astronomers were still calling them *planets*. To distinguish them from the larger planets, the terms *small planets*, *minor planets*, and *asteroids* were common in English publications, *asteroiden, planetoiden* and *kleinen planeten* in German[1], and *petites planetes* in French[2], but broad and consistent usage shows (until the mid 20th century) that these terms referred to a group of planets, not to a group of non-planets. A few examples of this usage are below:

---

[1] Later the compound *kleinplaneten* became popular in German, but this started mainly in the 1960s after they were no longer considered planets, as we argue below.
[2] English, German and French are the only three languages considered here.

… the sun is now well known to be placed in the centre, and to have **eleven primary planets** moving round him, each in its own path or orbit…The names of these planets, according to their distance from the centre of middle point of the sun, are, Mercury, Venus, the Earth, Mars, **Vesta**, **Juno**, **Pallas**, **Ceres**, Jupiter, Saturn, and Uranus… (Bonnycastle, 1816, bold added)

The first day of the present century is memorable in the annals of astronomy for the discovery of **the planet Ceres**; the first discovered of those telescopic bodies to which the term **Asteroids** has been applied. (Kirkwood, 1846, bold added).

The **small planets**, belonging to the extensive and remarkable group between Mars and Jupiter, have, by the common consent of astronomers, received the name of ***asteroids***. This term was originally proposed by the elder Herschel, and though perhaps open to criticism, has been so universally adopted, that it must now be regarded as their legitimate name. (Gould, 1849, bold added, italics in the original)

Observations, elements and ephemeris of the fifty-fourth **asteroid**…observed the following position of the **planet**…the ephemeris may be expected to represent the places of the **planet**, for a few weeks at least, with considerable accuracy. (Watson, 1858)

Professor C. H. F. Peters…has detected another **small planet**…This discovery is somewhat noteworthy, as completing the number **of one hundred known primary planets** in our solar system… But prior to the latter date the detection of the large group of **asteroids**, or (as they are now usually called) **minor planets**, had commenced. (Lynn 1867, bold added).

The four hundred **small planets** (**asteroids**) discovered up to 1898 all move in orbits situated entirely outside the orbit of Mars. (Campbell, 1901, bold added)

Only after the invention of the telescope, after understanding how to navigate the sky with it, did it become apparent that the number of **planets** [*Planeten*] was not limited to a few, but was rather great. Then the former older, big ones were distinguished from the newly discovered small ones as the **planetoids** [*Planetoiden*]. Almost 600 of them are already known in the sky,... (Oppenheim, 1911, trans. from German, bold added)

**The planets** [*Die Planeten*] are decomposed into:
1. inner planets [*innere Planeten*], medium-sized, very dense, faintly flattened. This subheading includes Mercury, Venus, Earth and Mars;
2. **small planets or planetoids** [*kleine Planeten oder Planetoiden*], more than half a thousand, all telescopic;
3. big planets [*grosse Planeten*], big, not very dense, very flat. Jupiter (the largest of all planets), Saturn (distinguished by its ring), Uranus and Neptune. (Kayser, 1920, trans. from German, bold added)

It is desirable that the list of **new planets** [*nouvelles planètes*] be published every year with detailed information regarding their discovery and with their elements…it is also to be desired that one publishes the ephemerides of **the four brightest minor planets** [*petites planètes*]**, and other planets** [*et d'autres planètes*] of particular interest. (IAU, 1922, translated from French, bold added)

Of the fourteen hundred-odd catalogued **asteroids** or **minor planets** revolving about the Sun, one of the most interesting sets constitutes what is commonly known as the Trojan group. The **asteroids**, it will be recalled, are **small planets** ranging in size from giant meteors, a few miles in diameter, to Ceres, with a diameter of some 480 miles. (Wyse, 1938, bold added)

The Commission recommends that final numbers be assigned to **new minor planets** [*nouvelles petites planètes*], when they have been observed at two oppositions and a satisfactory orbit was obtained. In the case of **a planet** [*d'une planète*] that passes closer to the Earth than the orbit of Mars, a definitive number can be given after a single opposition, provided that **the planet** [*la planète*] has been well observed, and that a satisfactory orbit has been obtained. (IAU, 1948, translated from French, bold added)

It has been assumed that the clusters would contract upon themselves, and that the final result would be, not a small number of large planets, but a large number of **small planets such as the asteroids**. (Edgeworth, 1949, bold added)

In this gap we find **thousands of small planets** called **asteroids**… (Whipple, 1968, bold added)

There are places in the literature where the word *planet* without qualification refers only to the larger planets, but it is normal in language for words to have both general and special meanings where the context makes clear which is being employed. In many cases an adjective was added to the word *planet* when only the larger or classical ones were intended: e.g., "the larger planets" (Newcomb, 1860), "the ancient planets" (Grove, 1866), "the principal planets" (Stockwell, 1873; Brouwer, 1935), "the great planets" (Chamberlin, 1901; Jeans, 1919;), "the well-known greater planets" (Newkirk, 1904), and "the chief planets" (Morrow, 1951). The need to do this indicates that asteroids were also a type of planet. The special standing that culture gave to the larger planets made it necessary to explain the smaller ones to non-specialists or to the public, such as the following published in *Science News-Letter* (now *Science News*) in 1951:

There are **thousands of known planets** circling our sun. Yet it is still quite right to say the chief planets are Mercury, Venus, our own earth, Mars, Jupiter, Saturn, Uranus, Neptune and Pluto, arranged in that order outward from the sun. **The other planets** are little bits of matter, ranging from several hundred miles across down to a city block. (Morrow, 1951, bold added)

On the other hand, the planetary science and astronomy community did not find it necessary to apologize or explain to one another when calling asteroids *planets*, which shows the taxonomy was widely understood in that community. Rather than allowing the modern planet concept to bias our interpretation of an earlier era, which is the well-known fallacy Perspective Bias, we should insist that the data shape our understanding of how *they* understood the planet concept. The data clearly show that the taxonomical concept of *planet* was allowed to stretch as ever-smaller bodies were discovered and admitted into the taxon, until the community had accepted hundreds and then thousands of very small planets. Below, we use the data in the literature to show when and why this suddenly changed.

## 2. <u>Too Many Planets</u>

The community began to feel overworked and perhaps even embarrassed by the large number of planets, but even this did not lead to any hints of reclassifying them:

> The Asteroid Problem…The rapid and continuous multiplication of discoveries…has introduced an embarrassment of riches which makes it difficult to decide what to do with them. Formerly the discovery of a new member of the solar system was applauded as a contribution to knowledge. Lately it has been considered almost a crime…so astronomical science has to determine whether it will go on with the work of discovery or drop the whole subject, as requiring a disproportionate amount of labor both for the observer and [human] computer…It must be admitted that there is a great deal of valuable time wasted in observing **minor planets**…Judging from our experience I should say that there must be **at least 1500 planets**, brighter than the fourteenth magnitude at some of their oppositions… (J.H. Metcalf, 1912, bold added)

> About three hundred had been discovered up to the year 1890 but discovery of new asteroids had become monotonous, and one well-known astronomer wrote of these small bodies as of "very little interest to anybody." (H.MacPherson, 1932)

> The discovery of **minor planets** has fairly been put on a basis of mass production. (H.N. Russell, 1932, bold added)

Throughout this period there was no discussion of making them non-planets, despite the burden of computing their ephemerides and the lack of scientific value the dynamical and statistical data provided at the time.

## 3. <u>They Sure Are Small!</u>

The thing that was continually noted that distinguishes asteroids from other planets was not their orbit sharing, but their small size, which was a geophysical rather than a dynamical feature:

> The [volume of the bodies] is a purely relative term, and it would be necessary, for the sake of consistency, to make three classifications out of **the nine primary planets**, since Mercury, Venus, Mars and Earth are very small compared to Jupiter. If Pallas is a hundred thousand times smaller than Mercury, then again, Mercury is many thousand times smaller than Jupiter. (Zach, 1802)

Of the **Telescopic planets**, Ceres, Pallas, Juno and Vesta…These four **planets** are so minute, that they can be only perceived by means of very powerful telescopes… (Laplace, 1830)

What is there improbable in the supposition that hundreds or even thousands of asteroids, too small to be detected by our telescopes, may revolve in this mysterious cluster? (Kirkwood, 1851)

It may not be remiss here to remark, for the consideration of those who regard the number of **small planets** as inexhaustible, the steady diminution in the size of these bodies as the progress of discovery advances. (Pogson, 1858, bold added)

Speaking of unknown **planets**, we are rather reluctantly obliged to admit that it is a part of our scientific duty as astronomers to continue to search for the remaining **asteroids**; at least, I suppose so, although the family has already become embarrassingly large. Still I think we are likely to learn as much about the constitution, genesis, and history of the solar system from these **little flying rocks** as from their larger relatives…Nor is it unlikely that some day the searcher for these **insignificant little vagabonds** may be rewarded by the discovery of some great world, as yet unknown, slow moving in the outer desolation beyond the remotest of the present planetary family." (Young, 1884, bold added)

And a very late example where asteroids are still called planets despite their small size, almost into the era when asteroids were reclassified as non-planets,

The sun's family of **planets** is divided according to size into three natural groups: the four major planets, Jupiter, Saturn, Uranus, and Neptune; the five terrestrial planets, Mercury, Venus, Earth, Mars and Pluto; and **the countless minor planets or asteroids**. The asteroids are **so small** relative to the **other planets** that their admission to planetary rank is generally qualified by reservations that are almost apologetic. (Nicholson, 1941, bold added)

## 4.  <u>They Are Everywhere!</u>

There was a growing recognition that these small bodies, while primarily in the band between Mars and Jupiter, actually share orbits with all the planets of the solar system and are probably the source of meteorites because Earth is not excepted from sharing orbits.

Furthermore, the large number of asteroids now known to exist and their rapidly increasing number, give us ground for believing that an innumerable quantity of infinitely smaller bodies may be revolving in space and when they come sufficiently within the attraction of earth, they are precipitated to its surface. (Harris, 1859)

The number of known asteroids, or bodies of a smaller size than what are termed the ancient planets, has been so increased by numerous discoveries, that instead of seven we now count **eighty-eight as the number of recognized planets**…If we add these, the smallest of which is only three or four miles in diameter, indeed cannot be accurately measured, and if we were to apply the same scrutiny to other parts of the heavens as has been applied to the zone between Mars and Jupiter, it is no far-fetched speculation to suppose that between these asteroids and the meteorites, bodies of intermediate size exist until the space occupied by our solar system becomes filled up with planetary bodies varying in size from that of Jupiter (1240 times larger in volume than the earth) to that of a cannon-ball or even a pistol-bullet…another half century may, and not improbably will, enable us to ascertain that the now seemingly vacant interplanetary spaces are occupied by smaller bodies which have hitherto escaped observation, just as the asteroids had until the time of Olbers and Piazzi. (Grove, 1866, bold added)

The modern orbit-clearing theory has a response for this, arguing that some small degree of orbit-sharing does not negate the primary fact of gravitational dominance by the larger bodies. The theory that brings this response into focus was developed at a much later time than the era we are investigating, and we find no evidence of this argument from that era. Projecting the orbit-clearing motive back onto that era is anachronistic.

## 5.  Were Useful as "Planets"

During this period, the taxonomical identification of asteroids as planets was scientifically useful and supported hypotheses that helped their understanding of the solar system. In the scientific process, taxonomy organizes our observation of phenomena into categories so we can think and speak clearly about the patterns as we develop hypotheses and models. Scientists thus evolve taxonomies to be useful for hypothesis-formation and model-building, and this is what drives definitions like *planet*.[3] In a future manuscript (in preparation) we discuss this in the larger context of planetary science since the Copernican Revolution. Regarding asteroids, throughout the nineteenth and earlier twentieth century, scientists had no motive to create taxonomies in which asteroids were non-planets because they were not yet working on hypotheses or developing models that explain the differences between asteroids and other planets. In fact, keeping asteroids and planets in the same taxonomical category was scientifically useful for the models that *were* being discussed during that era. Since asteroids were seen as just smaller planets that underwent the same formation and evolution as larger planets, scientists could infer from fragments of asteroids (the meteorites) the composition of the inaccessible interiors of their siblings, the terrestrial planets.

> And from the chemical composition of these wanderers, we can justly infer, that other worlds are composed of the same simple elements, and that the same chemical and physical laws prevail throughout the universe. (Harris, 1859, p. 13)

Laplace's Nebula Hypothesis (Laplace, 1829) was the idea that a primordial nebula contracted to become the sun, shedding successive rings of material from its equator to conserve angular momentum during the contraction, each ring then condensing into planets with their secondary systems of satellites and rings in similar manner. The asteroid belt came to be viewed as a ring where the process of condensing and consolidating was incomplete, leaving many small planets instead of one large planet. Because all planets must have formed by the same condensing processes (this was the hypothesis flowing from the taxonomy), they could infer the same processes operated on the larger planets where the evidences are now swept away. Several astronomers worked this way beginning from a binary taxonomy of planets vs. comets, where the asteroids were included in the former category:

> Besides the central or controlling orb, [the solar system] contains, so far as known at present, **sixty-seven primary planets**, twenty-one satellites, three planetary rings, and nearly eight hundred comets…The *cometary*, is distinguished from the *planetary* portion of the system by several striking characteristics…(Kirkwood, 1860, bold added, italics in the original)

---

[3] This is why taxonomical freedom is vital for the progress of science.

Because the asteroids shared the relevant characteristics of prograde motion with relatively low inclination and eccentricity compared to comets, they fit into the taxonomy as planets and were evidence supporting the model. Both Kirkwood and Giovanni Schiaparelli used this taxonomy to argue comets do not share the angular momentum of the solar nebula from which the planets (including asteroids) formed, and therefore comets are from interstellar space. This extended the explanatory power of the model. Schiaparelli's views were described by Lassell:

> **This system seems to consist of two classes—the Planets**, characterized by but little eccentricity of orbit, slight variation in the plane of the orbit, exclusion of retrograde motion, and a tendency to take the form of a sphere[4] (deviating from it only so much as is necessary to preserve the equilibrium of the body)—these characteristics applying also to the secondary systems [moons], with the exception of the satellites of *Uranus*. The second class consists of **cometary bodies**, which are under no law as to the planes of their orbits, or the direction of their motions. The point most remarkable about them is the extreme elongation of their orbits, most of which are described in interstellar space; which seems to show that they did not form part of our system when that was first constituted, but are wandering nebulæ picked up by our sun. (Lassell, 1872, bold added).

Kirkwood further extended the explanatory power of the Nebula Hypothesis by addressing the unique characteristics of the asteroids as a special case of planets, thereby strengthening the hypothesis, itself:

> These bodies occupy a chasm—observed before their discovery—in the order of the planetary distances; an order which indicated the existence of **a single planet where fifty-eight** have already been detected…
>
> The members of the group are characterized by certain peculiarities which are doubtless indicative of an intimate mutual relationship. They are extremely diminutive in size…The

---

[4] The non-spherical shape of most asteroids was unknown at the time and there was a bias, discussed below, in assuming they are spherical.

eccentricities and inclinations of their orbits are generally much greater than those of **other planets**. Their orbits interlace. "The strongest evidence," says D'Arrest, "of the intimate connection of **the whole group of small planets** appears to be, that if the orbits are supposed to be represented materially as hoops, they all hang together in such a manner that the entire group may be suspended by any given one." (Kirkwood, 1860, bold added)

Kirkwood argued from this relatedness of the asteroids that they formed from one of the rings of material shed by the contracting solar nebula, which, unlike the other rings, has continued as a group of small planets rather than condensing into a single large planet due to the stabilizing actions of Mars and Jupiter. Kirkwood developed this view by analogy with the rings of Saturn and its moons:

It has been affirmed by an eminent astronomer that the rings of Saturn are sustained by the direct action of the satellites; that no other planet of the system has such an arrangement of secondaries[5] as to secure the stability of a ring; and that the only place in the solar system where a primary ring would be long sustained, is the region occupied by **the small planets**…

Now according to the nebular hypothesis each of the principal planets originally existed as a gaseous ring. The observed order would require such an annulus between Mars and Jupiter, and this is the precise situation in which a ring would be the longest sustained by the exterior planets. Upon the breaking up of the ring, however, *a zone of **small planets** would naturally be formed* unless some one portion of the vaporous mass should have a preponderating influence so as ultimately to absorb all the rest. In short, it is believed that the nebular hypothesis will explain the various phenomena of these bodies, which without it seem inexplicable… (Kirkwood, 1860, bold added, italics in the original)

Kirkwood (1869) later argued from the Kirkwood Gaps in comparison to resonances in the Saturn system that nucleation of planetary cores and subsequent growth in this ring would be governed by the disturbance of Jupiter. Therefore, both the minor planets and Saturn's rings "furnish strong arguments in favour of the nebular hypothesis" and are "further suggestive as to the *mode* of planetary formation." (Kirkwood, 1869, italics in the original)

To summarize, scientists in this era called asteroids *planets* because the taxonomy was aligned with the Nebular Hypotheses in which planets of all sizes shared a common formation process albeit with perturbations that allowed evidence of the process to survive among the smaller planets. The hypothesis therefore became stronger, not weaker, as many more of these small planets were discovered, showing ever more evidence of the planet formation processes. This point was noticed and stated by a contemporary:

---

[5] I.e., secondary planets (satellites), so-named as a sub-class of planets since Kepler. Sun-orbiting planets are sometimes called *primary planets* or *primaries* even into the present. This will be discussed in detail in a future manuscript (in preparation).

The double fact of the **multitude** of those bodies and of their circulation in the same direction around the sun, is now too imposing to admit an explanation of their origin and formation different from that developed by Laplace…" (Plana, 1856, bold added)

Thus, the growing number of asteroids strengthened the scientific usefulness of their taxonomical identity as planets; this is precisely opposite the modern, incorrect claim that a growing number of asteroids motivated astronomers to reclassify them as non-planets (which finds no support in the literature). Eventually this method of reasoning no longer worked when the available insights were exhausted and the asteroids were found to be geophysically very different than the planets.

## 6.  Well, They Aren't All Planets

As planet formation theories developed, geophysical reasons were discovered to think of asteroids as different than planets, so taxonomies developed to reflect the new thinking.[6] The beginning of this change is found in a paper by Kuiper in 1950. He recognized from photometry that many asteroids had non-spherical shapes, and the growing connection with meteorites suggested their fragmentary nature, so he created a planet formation hypothesis that incorporated these observations (Kuiper, 1950). He calculated that "5-10 planets" could form by condensation between Mars and Jupiter, then collisions would break most of them apart into smaller fragments. He said,

> The largest asteroids, like Ceres, are regarded as such original condensations; they are in effect **true planets**. (Kuiper, 1950, bold added)

> The total number of small proto-planets estimated to have formed in the region between Mars and Jupiter is between 5 and 10. They formed **small planets, like Ceres**…It is assumed that two of these collided sometime during the last $3.10^9$ years, an event having a sufficiently large probability. Thereafter secondary collisions became increasingly frequent. The most recent of these collisions account for the Hirayama families. In this manner **thousands of asteroids were formed, being the largest of the fragments**, as well as billions of meteorites. (Kuiper, 1951, bold added)

He does not use the term *planet* for the smaller asteroids because they are but fragments of the original planets, and the even-smaller fragments are meteorites. This is the first widely accepted taxonomy in which the bodies of the asteroid belt are divided into two fundamentally different categories: planets versus asteroids (or three, counting meteorites). In Kuiper's scheme it was not the smaller asteroid's existence in a cluster that made them non-planets, but their intrinsic existence as broken fragments regardless where they were or how many neighbors they had. The larger bodies like Ceres were still planets although they shared these orbits.

---

[6] An early hint of what would come in the 20th century may be found in R.P. Greg's argument from meteoritics (Greg, 1854) in favor of Olber's hypothesis (Cunningham, 2017) that asteroids are the fragments of a disrupted planet. Greg tended to avoid calling asteroids *planets* except when quoting others, but he did call them *fragmentary planets*, suggesting they are a particular subset of planets. Note this was a geophysical taxonomy based on intrinsic properties of the bodies as fragments versus whole planets, not a dynamical taxonomy based on the orbits these fragments attained. By Greg's time, astronomers had generally discarded Olber's hypothesis (Kirkwood, 1869), so Greg's argument was an outlier.

Later, Kuiper examined whether gravitational instability was adequate to produce the asteroid belt:

> This clarification of the satellite problem suggests a re-examination of the planet problem, in particular that of the **planets of small mass, the asteroids**. The hypothesis made before was based on the assumption that gravitational instability, clearly operative in the formation of the 8 major planets, was also responsible, though in modified form, for the formation of **the original group of minor planets**…it appears doubtful to the writer now that the anticipated development could arise from gravitational instability alone…It has become clear that uniformity[7] does not exist in satellite formation and it appears necessary to abandon it also between planets and asteroids. (Kuiper, 1953, bold added)

Beginning from this point in the paper his choice of terms reflects a new taxonomy with two categories, the "8 major planets" that formed by gravitational instability, and the asteroids that formed by accretion plus collisional fragmentation:

> On the accretion hypothesis the number of original asteroids cannot be clearly determined, but it is expected to be rather large, say 10-100 in the observable region. Collisions within the asteroid ring will have started much earlier, and will have been more frequent later on…Furthermore, not all small asteroids need be fragments of collisions; on the old hypothesis the production of small asteroids from protoplanets would have been even more artificial than of Ceres. (Kuiper, 1953)

In other words, prior to understanding the role of accretion, it was incredible to imagine bodies smaller than Ceres forming directly by condensation (as very small, gaseous protoplanets) so they were explained as exclusively collisional fragments; however, the new accretion hypothesis accounts for small bodies forming directly, although many of the small bodies would still be collisional fragments. Kuiper called the later, mixed set of undisrupted bodies and fragments *asteroids* without distinction. Fred Singer (1954), however, distinguished among the evolved set of bodies, calling the surviving accreted ones *planets* and *proto-planets*, but the fragmented pieces *asteroids*. Thus, Ceres and Vesta (for example) would be planets in Singer's taxonomy while fragmented asteroids would not be. These theoretical developments set the stage for what happened next, at the end of this era, by providing a framework in which asteroids can be understood as geophysically different than planets.

## 7. <u>Flood of Geophysical Data</u>

Finally, the literature shows that a flood of geophysical data about asteroids arrived in the later mid-20th century – judging by the widening range of topics in published papers, it seems to have started around the 1960s – and this is precisely when the terms *minor planet* and *small planet* suddenly died out from published usage in favor of calling them only *asteroids*. To test this correlation, we collected data from Google Scholar searches on the year-by-year numbers of scholarly papers that included these terms.

---

[7] I.e., uniformity of formation process, all bodies forming by the same process.

Figure 1 shows that the overall number of published planetary science papers has been growing exponentially, and all relevant terms likewise grow throughout the period. This represents the extensive growth of the planetary science community but does not reflect the intensive views on terminology it employed.

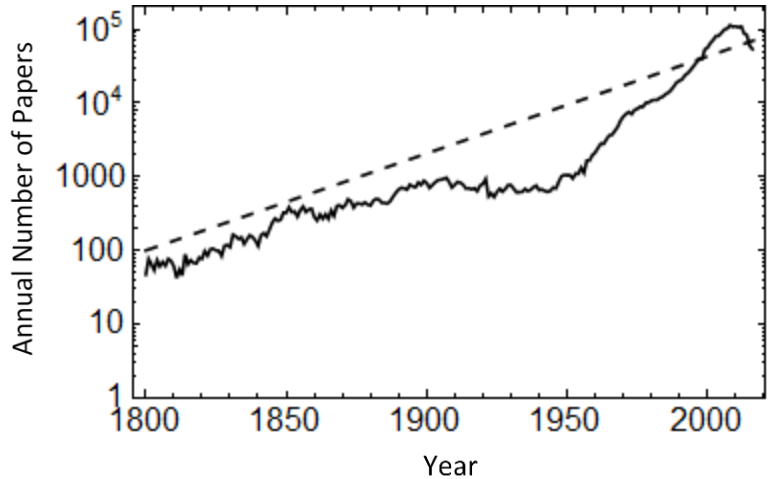

**Figure 1.** Yearly number of scholarly papers in all languages that include the word "planet". Dashed line: exponential trend with 3.08% annual increase.

Fig.2 shows the number of yearly papers that include the terms *small planet* and *minor planet* as a percentage of the number of yearly papers that included the term *planet*. It shows that *small planet* was a popular term as soon as Ceres was discovered, and its relative use grew yearly until the term *minor planet* became similarly popular. The small rise at the end of the graph for *small planet* represents usage of the term for Kuiper Belt Objects, not asteroids. The term *minor planet* eventually displaced *small planet* then grew to its heyday in the 1950s, just after the founding of the Minor Planet Center in 1947. It rapidly fell out of favor in the late 1950s as *asteroid* became the only common term. The continued rise in uses at the end of the plot for *minor planet* is mostly from "Minor Planet Circular" and "Minor Planet Bulletin" appearing in the reference sections of manuscripts, especially as KBOs are discovered beginning in 1992.

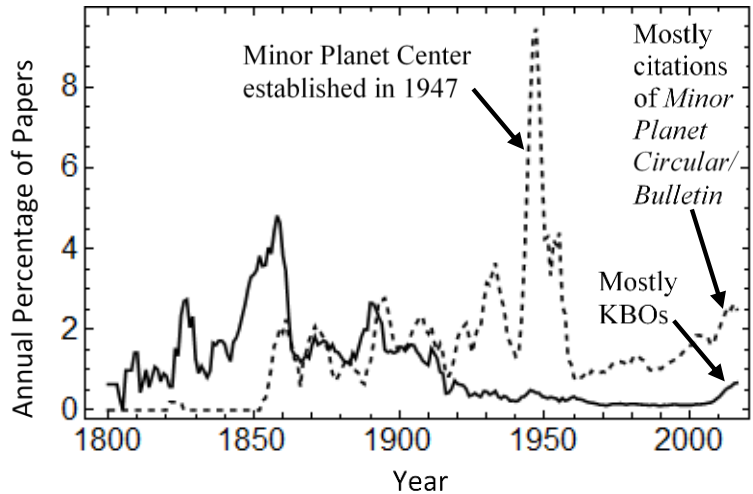

**Figure 2.** Yearly percentage of the scholarly "planet" papers that include the term "small planet" (solid) and "minor planet" (dashed), shown with five-year smoothing for clarity.

Fig. 3 shows the normalized use of *asteroid*. It has always remained popular, and papers including the term have constituted around 10% of planetary science papers since the mid-19[th] century, with recent dramatic growth to nearly 20%. Its popularity throughout that time does not imply asteroids were considered non-planets. Instead, the term was used to distinguish the asteroids as a particular subset of planets, as the literature review clearly showed.

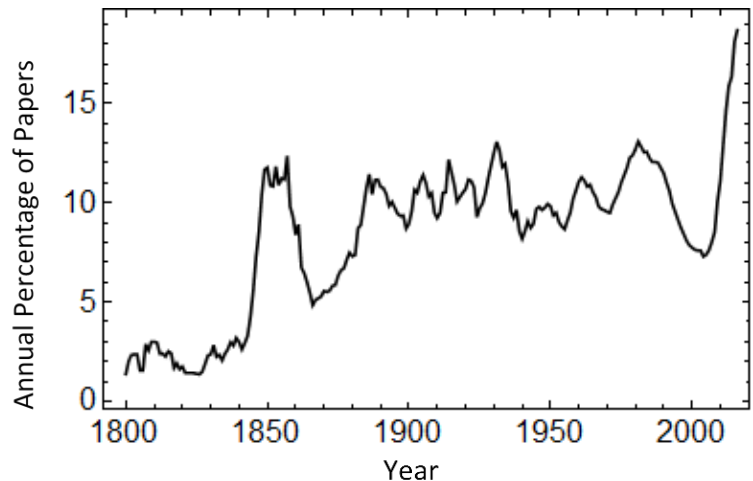

**Figure 3.** Scholarly papers that include the term "asteroid" expressed as a percentage of scholarly papers that include the term "planet", shown with five-year smoothing for clarity.

Figure 4 shows the ratio of papers containing either *small planet* or *minor planet* to the number of papers containing *asteroid*. The ratio averaged about 44% from 1801 until 1957. There was no drop in usage of the two *planet* terms in the latter 1800s as many asteroids were being discovered, which demonstrates orbit sharing was not a taxonomical consideration. There was a small reduction in usage in the 1900s followed by a sharp revival when the Minor Planet Center (MPC) was established. However, despite the existence of the MPC, there was a sudden and dramatic drop from 1956 to 1961 to a floor of 11%, which it held for the next three and a half decades demonstrating that a real change in attitudes had suddenly occurred. Beginning about 1995 there was increased use of the term "minor planet", corresponding to the discovery of Kuiper Belt Objects (KBOs) beginning in 1992. We checked that this rise was largely due to increasing citations of the "Minor Planet Circular" and "Minor Planet Bulletin" in reference sections of papers, including reports on the rapidly accelerating discovery of KBOs. These data show that the only significant shift in terminology occurred in the 1950s, correlating to the influx of new geophysical data.

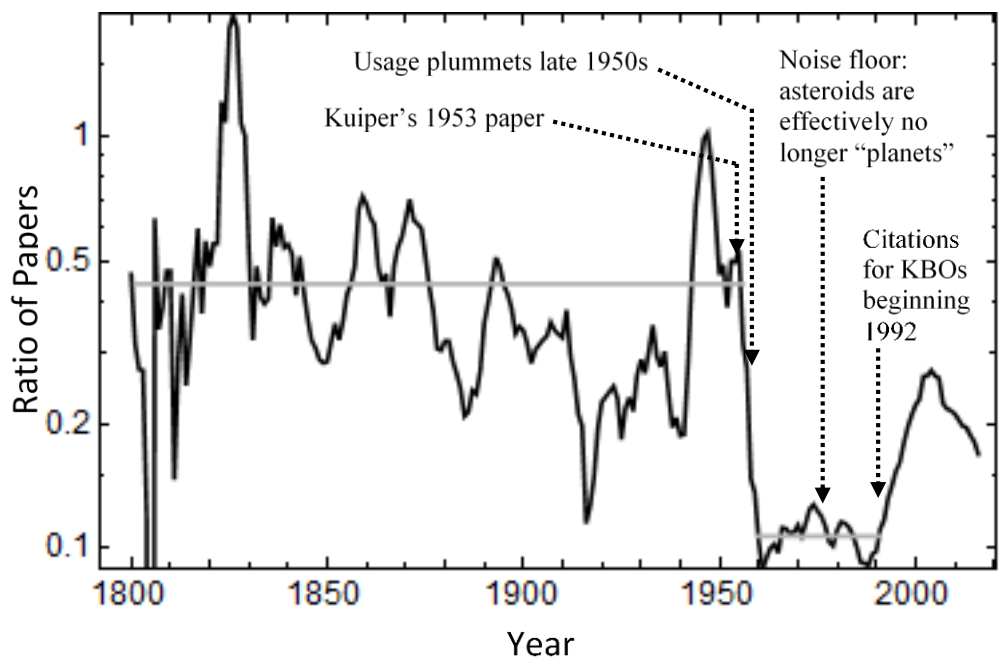

**Figure 4.** Semilog plot of ratio of the number of scholarly papers containing either "small planet" or "minor planet" to the number of scholarly papers containing "asteroid", with 5-year smoothing for clarity. Horizontal gray lines: guides to the eye for average values over 1801-1956 and over 1960-1991.

We checked these results by investigating the use of terminology in specifically German and French publications. For the German language, we found the number of publications each year that contain any one or more of these terms: "*kleinen Planete*", "*kleinen Planeten*", "*kleine Planeten*", "*Asteroide*", "*Asteroiden*", "*Planetoide*", "*Planetoiden*", which we call set A. These are used in papers about small bodies regardless whether they consider them to be planets or not. (Note that the Google Scholar searches did not distinguish capitalization.) We also found the number of publications each year that contain any one or more of these terms: "*kleinen Planete*", "*kleinen Planeten*", or "*kleine Planeten*", which we call set B. These represent the terms for small bodies that literally call the bodies planets, although the usage may not intend to convey that meaning. Late German usage also contains the compound term "*Kleinplaneten*". We found this was almost never used before about 1960. We found it used in papers where asteroids are treated as non-planets. It was never used significantly in the period where asteroids were treated as planets. Qualitatively we found that many papers prior to about 1950 use all the terms in set A as a subset of the *Planeten* [planets], implying that they were considered planets regardless which term was used. Naively we might think use of the term *Planetoiden* indicates the author considered the bodies to be non-planets, since the root word for planet had been modified with the suffix *-oid*. However, we found this term was in fact used to mean planets in many verifiable cases, and we never found any cases that explicitly show it was meant to refer to non-planets. After about 1950 we did not find any more papers explicitly calling the small bodies *Planeten*, suggesting a change took place somewhere close in time to the shift in English usage. Quantitatively, Fig. 5 shows the ratio of set B to set A. The set B terms began dropping in usage from about 80% of all small body papers prior to 1950 to only about 20% by the late 1960s. This timing correlates well to the growing popularity of the term *Kleinplaneten* starting in the 1960s, which converted the set B terms into a compound noun with a technical meaning that, judging by usage, did not imply the bodies are planets. The timing of all these changes correlate well to the shift found in the usage in English language papers.

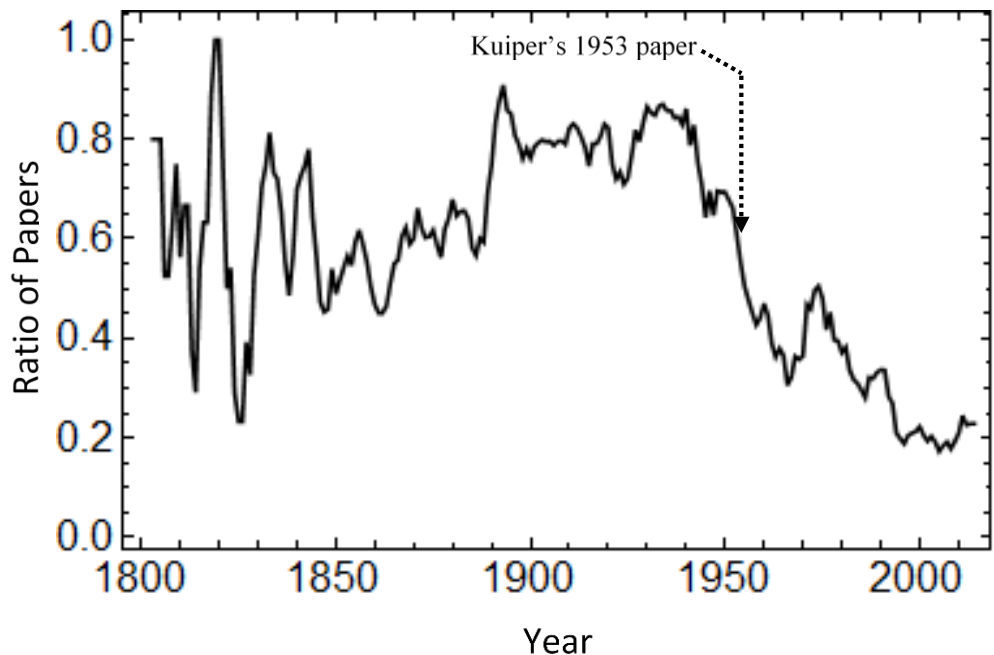

**Figure 5.** Plot of ratio of the number of German-language scholarly papers containing either terms "*kleinen planete*", "*kleinen planeten*", or "*kleine planeten*", without regard for capitalization, divided by the number using any of these terms or "*asteroide*", "*asteroiden*", "*planetoide*", or "*planetoiden*" without regard for capitalization, with 5-year smoothing for clarity. The drop in this ratio occurred at the same time the change occurred in English language terminology.

For the French language, we studied the number of papers each year that used one or more of the terms *petite planète*, *petites planètes*, *asteroide*, or *asteroides*. We found qualitatively that, just as in English and German, there are many uses of all these terms up until about 1950 where it is clear they are considered a type of planet, but after about 1950 we found no more such cases in our sampling. Corresponding to this, and paralleling the English and German usage, the terms *petite planète* and *petites planètes* began losing popularity beginning about 1950. However, in the 1970s these French terms had a revival, then dropped away again into relative disuse by about 2000. We suspect the slower abandonment of *petite planète* in French than of *small/minor planet* in English and *kleinen Planeten* in German was because the French language had no ready-made alternative term. The literature shows that the popular German term, *planetoid*e, was almost never used in French, while the popular English term, *asteroid*, was unpopular in French publications (as *asteroide*) throughout the 19[th] century, starting to grow in usage only after 1900 and then slowly. Thus it may have been easier in French publications for *petite planète* to take on the technical meaning that no longer implies planethood than to switch to one of these terms. By contrast, in English and German the unintended implication of planethood could be avoided more unambiguously by switching to the existing, popular term *asteroid*, *Asteroide,* or *Planetoide*.

These changes in terminology in English, German and French correlate well to the revolutionary influx of geophysical data about asteroids. Prior to that influx, the true character of asteroids was unknown:

> As regards the physical features of the asteroids, we at present know practically nothing…I am persuaded that a knowledge of the substance, form, density, rotation, temperature, and other physical characteristics, of one of these little orphans, would throw vivid light on the nature and behavior of interplanetary space, and would be of great use in establishing the physical theory of the solar system. (Young, 1884)

> The **asteroids** are individual **planets** possessing physical characteristics just as the Earth and Moon, but they are so small that no direct observations can be made of their surfaces. (Nicholson, 1930, bold added)

That lack of information, coupled with inadequate formation theory, allowed scientists to imagine asteroids as more planet-like than they actually are, and this may have contributed to the long willingness to keep them classified as planets. For example, some of the literature through the early 1900s suggested an unwritten assumption the minor planets are spherical. (Not all believed this; as early as 1802 Zach wrote, "That Ceres and Pallas are both of very changeable luminosity – this I explain in that both planetary fragments are not round, but rather of a very irregular shape" [Zach, 1802].) Laplace (1829) had argued that planets condense in spherical form. Kirkwood (1860), after counting sixty-seven primary planets, wrote, "The spheroidal figure of the planets points to a great and significant fact in regard to their primitive constitution—the fact that they have *all*, at former epochs in their history, been in a liquid, if not in a gaseous state" (italics in the original). Regarding Mars' oblateness he wrote, "On the nebular hypothesis we have only to suppose that in the process of transition from the gaseous to the liquid and solid form, 'the liquid surface of some planets was solidified before they could assume the figure appertaining to their velocity of rotation'," attributing the quote to Humboldt (1860). We found no indication that planetary formation theories of that era had begun to predict non-spherical primitive bodies. That development came later. The light curve of Eros suggested to multiple observers that it was a "double planet", or "la Planète double (433) Eros", consisting of two co-rotating spherical or ellipsoidal bodies rather than a lumpy, single body (André, 1901; den van Bos and Finsen, 1931; Pickering, 1932). The spherical assumption was eventually overthrown by thinking hard about the photometric curves, but the assumption ensured this did not happen immediately:

> There is no doubt that minor planets show variations in their continuous spectrum. This can be reasonably explained as an effect of rotation and of different reflectivity of the portions of their surface. (Bobrovnikoff, 1929)

> Several of the asteroids undergo small but uniform periodic changes in light in less than one-half a day, a variation which can be attributed to rotation and spottedness. But Eros has shown remarkable variations in range, as well as a possible change of period, which cannot be explained so simply. The only hope of solving the mystery lies in obtaining a long series of systematic observations of its magnitude. (Harwood, 1930)

> Some of the minor planets are irregular in shape as indicated by the variations in the light reflected as they rotate. This suggests that they are the broken fragments from a collision of objects of more uniform surface brightness and spherical shape such as we expected as the result of almost any accretion mechanism. (Urey, 1951)

Finally, in the 1960s the literature appears to change qualitatively, which suggests this begins the "modern era" of asteroid studies. We begin seeing asteroid shape modeling studies, which furthered the picture that asteroids are not planet-like. There were many studies of collisions in the asteroid belt, also suggesting these bodies are not planets but instead fragments of (proto)planets. Spectroscopic studies abounded, identifying mineralogy, with many geophysical analyses of asteroids, not just the dynamical studies of the prior decades. Asteroid taxonomies begin appearing, but they did not call the asteroids planets and the taxonomies focused on spectral classes and collisional families. (e.g., Chapman et al., 1975; Bowell et al., 1978; Tholen, 1984; Barucci et al., 1987; Tholen and Barucci, 1989). Another shift in perception arose with the determination that asteroids were themselves accumulations of fragments (Chapman et al., 1977; Davis et al., 1979), resulting in their being referred to as "rubble piles" in the literature for the first time in 1979 (Davis et al., 1979). The concept of asteroids had evolved very far from the classical concept of planets. Altogether, this history indicates that it was geophysical characteristics, not sharing of orbits, that led to the shift in terminology in which asteroids were no longer called planets.

As the ever-smaller asteroids were admitted into the *planet* taxon, the concept of *planet* itself was effectively – and to some degree unwittingly – stretched or distorted, because the small asteroids were geophysically dissimilar to other members of the taxon. Their removal from the taxon effectively restored the meaning of *planet* to what it was before their discovery. In the restored taxonomy, Ceres might still be included as a planet as Singer suggested. Scientists have in fact continued calling Ceres a planet and arguing it should be classified as such based on its geophysical complexity, and increasingly so as the Dawn mission has sent back detailed geophysical data:

> …all of the alternatives…are in difficulty if one is required to maintain the assumption that **the planet's** composition is essentially uniform throughout. The writer would like to propose a different model for **Ceres** which would keep some or all of the interpretations of surface materials as live possibilities, and yet explain the low density of **the planet**…The writer is uncertain whether the ice of the mantle region also extends down into the core of **the planet**. (Hodgson, 1977, bold added)

> "**Planet Ceres**… Ceres is also unlikely to be genetically different from other planets: it, too, is the end product of an accretion process from the early solar nebula… Let us accept Ceres, the largest planetoid, as **one of the planets** – as far as that makes sense: in considerations of planetary orbits and the structure of the solar system, and also in the context of comparative planetology and planetary cosmology." (Hantzsche, 1996, translated from German, bold added)

Ceres orbits the sun and is large enough (1000 km diameter) to have experienced many of the processes normally associated with planetary evolution. Therefore, it **should be called a planet** even though it orbits in the middle of the asteroid belt. (Sotin, 2003, bold added)

Ceres orbits the Sun and is large enough in size to have experienced many of the processes normally associated with planetary evolution. Therefore it **should be called a planet**. Its location, in the middle of the asteroid belt, has caused it to be referred to as an asteroid. However, its size, orbit and general nature, as best we can discern it, suggest that it is much more interesting than the small pieces of larger objects (perhaps such as Ceres) that one normally thinks of when one refers to asteroids. **Ceres' planet-like nature** and its survival from the earliest stages of solar system formation, when its sister/brother objects probably became the major building blocks of the Earth and the other terrestrial inner planets, makes it an extremely important object for understanding the early stages of the solar system as well as **basic planetary processes**. (McCord and Sotin, 2005, bold added)

Ceres likely contains considerable water, has differentiated, and has experienced major dimensional and chemical changes over its history, making it **more a planet than asteroid**. (McCord, 2013, bold added)

…the Exploration of Ceres by Dawn...The first comprehensive survey of **the planet** is scheduled to commence in late April 2015… (Russell, et al., 2015, bold added)

The spikes were consecutively located closer to **the planet** as the spacecraft approached the subsolar point. The dramatic increase in electron counts could be explained by the acceleration of the electrons at a Ceres bow shock. (Villarreal et al., 2015, bold added)

Third-Order Development of Shape, Gravity, and Moment of Inertia of Ceres …interpreting shape and gravity data in terms of interior structure and infer deviations from hydrostaticity that can bring information on the thermal and chemical history of **the planet**. (Rambaux et al., 2015, bold added)

To a lesser extent, Pallas and Vesta have been called planets. They are protoplanets (McCord and Sotin, 2005), which literally means "first planets" (protero- mean pre-, while proto-means first). Hodgson (1977) called all three "planets". McCord and Sotin (2005) argued that Ceres displayed more active planetary processes than the other two and that water content made the difference. The Dawn mission showed Ceres and Vesta are both geologically active (Hoffman et al, 2016). Whether borderline cases are planets or not is the sort of taxonomical argument that leads to greater clarity in taxonomy.

The great influx of geophysical data on all the other, smaller asteroids made it clear they are so unlike planets such as the Earth and the Moon that it became no longer scientifically useful in hypothesis-formation to treat them as the same category of body. Instead, the vast, geophysical differences between asteroids and (what we now call) planets made it more scientifically useful to classify them as different types of bodies and thus form hypotheses about the processes that made them different. The changing landscape of scientific usefulness based on the availability of geophysical data is what drove the reclassification.

## Discussion

It is untenable to believe asteroids were reclassified as non-planets on the basis of orbit sharing or large numbers, not just because these kinds of arguments are severely lacking or absent in the literature, and not just because we can see scientists continuing to call them *planets* as the number of orbit-sharers grew to huge proportions, but because the development of orbit clearing theories came at a later time in history; it is a modern concern and so it is anachronistic to insist an earlier era shaped its taxonomies to promote it.

The one exception mentioned at the beginning of this discussion was the 1802 paper by W. Herschel, where he did argue for non-sharing of orbits and had a theory that substituted for the modern orbit-clearing arguments; his theory was the Titius-Bode Law. He hypothesized asteroids must be a different class of body than planets because otherwise it would upset the harmony of Titius-Bode. Apparently, his contemporaries did not find this compelling because they immediately dropped it; perhaps this is because Titius-Bode would still be a failed theory if there were no planet in the slot between Mars and Jupiter, so either way the harmony would be broken, or perhaps a group of small planets was felt to play an adequate role maintaining the harmony of Titius-Bode. Either way, the fact that the Titius-Bode Law was not a true physical theory probably made it easier to set aside. After Herschel's argument was abandoned, there was never again a claim that the non-sharing of orbits should be in the definition of planets until the 21[st] century when the IAU needed to decide who would name the many planet-sized bodies being discovered in the Kuiper Belt. This is when the spurious argument appeared that "just as asteroids are not planets because they are in a belt, so also KBOs should not be planets because they are in a belt." As we have shown, there is no historical support for the premise of that argument. By the time of the IAU dispute, consensus had indeed formed that asteroids are non-planets, but only very recently and only on the basis of their geophysical properties rather than their dynamical status. The basis for that consensus does not apply to the larger KBOs just as it does not apply to Ceres.

The transition in the literature away from calling asteroids *planets* appears abruptly. We found this not only in the plot of data in Figure 4, but also in our attempts to locate key papers in the mainstream development of planet formation theory calling asteroids *planets*. These were easy to find until a certain point in history, then suddenly extremely difficult. The change in the English language appears to us as a first-order phase transition (a step function), not a mere second order phase transition (a change in slope). What can explain such an abrupt transition? Prior to Kuiper's 1953 paper quoted above, key publications in the mainstream of planet formation theory freely called asteroids *planets*. Kuiper himself did so, but then he changed terminology within the text of his 1953 paper, saying "planets of small mass" and "minor planets" when describing the prior theory, switching exclusively to "asteroids" in his new theory. We found none of the key papers on planet formation immediately citing Kuiper (1953) calling them *planets*, and very few of any key papers after Kuiper (1953) that did (apart from papers about Ceres). Kuiper seems to have provided the vital theoretical step that changed the language of theorists and that adequately explained what was about to be seen in the flood of geophysical data. If we wish to pin the reclassification of asteroids on a specific, paradigm-shifting event in history, the evidence suggests that Kuiper's publication in 1953 would be that event.

This taxonomical development is identical to what happened in the Copernican Revolution, when Copernicus provided a new theoretical framework then a new influx of geophysical data arrived as Galileo pointed his telescope at the Moon, revealing it to be Earth-like with mountains (Galileo, 1610). The observation of an Earth-like Moon was vital in the debate over heliocentrism (Fabbri, 2012; 2016), because if planets are Earth-like worlds, not idealized spheres with a heavenly physics, then it is a short leap to conclude the Earth is a planet in the heavens as much as they are. Thus, Galileo and other astronomers immediately adopted a new taxonomy where *planet* became a geophysical classification of bodies, the Earth-like worlds including the Moon but no longer the Sun. This replaced the old taxonomy where planets were a dynamical class of lights in the sky. In fact, in the new Galilean-Copernican taxonomy the planets no longer shared a common dynamical state: some orbited the sun, while others—like the Moon that orbits the Earth and the "Medicean Planets" that orbit Jupiter (Galileo, 1610)—did not orbit the sun. Thus, it was consideration of scientific usefulness in promoting geophysical theories about nature—not the mere dynamical status of bodies as a cultural affair—that drove the re-definition of *planet*, both during the Copernican Revolution and when reclassifying asteroids. In both cases, a new taxonomy was chosen to align with both the available geophysical data and the reductionist geophysical theory that sought to explain the data. The taxonomy provided the new conceptual categories and corresponding language that scientists needed to advance their hypotheses. This is argued further in our future paper.

The broad consensus that asteroids are not planets (apart from Ceres as a protoplanetary survivor of sufficient size to be geologically complex) was established without voting by any professional body such as the IAU, because adequate time (160 years) was allowed for the full scientific community to freely come to consensus. This process was carried out the usual way in peer-reviewed journals and in scientific conferences. Similarly, for more than 150 years the biologists have vigorously debated many possible and irreconcilable definitions for the taxonomical concept *species* (see, e.g., Hey, 2006; Zinner and Roos, 2014; Aldhebiani, 2018), one of the most important concepts to biology just as *planet* is to planetary science, yet their professional bodies have refrained from attempting to force closure by a vote. Voting on key taxonomical terminology and the relationships between taxa is anathema in science (see e.g., Ride, et al., 2012) and is contrary to the traditions evolved over many centuries to reduce social, political and personal cognitive biases in science. It injects unhelpful dynamics and social pressures into science and impinges on individual scientists' taxonomical freedom. Taxonomical freedom for the *planet* concept is especially important as new exoplanetary and transneptunian data are being generated.

## Conclusion

We were unable to find any cases after Herschel's 1802 paper where scientists argued that asteroids are not planets on the basis of orbit-sharing, until the idea was anachronistically reflected backwards during the 2006 IAU controversy. Instead, the literature gives strong evidence of several other trends:

1. Scientists pervasively considered asteroids to be a sub-classification of planets throughout the 19[th] and early 20[th] centuries until the mid- to late-1950s.
2. Scientists freely admitted hundreds of planets into the *planet* taxon and expected there would be thousands. The work of discovery and computing orbits became a burden of questionable value but this did not lead them to question the taxonomy.
3. The concept of *planet* was allowed to evolve to include ever smaller bodies as these discoveries continued, yet they still did not question the taxonomy.
4. It was realized that these ever-smaller bodies exist throughout the solar system, and they connected them to the meteorites that fall on Earth.
5. Planet formation theories benefitted from the taxonomical identification of asteroids with larger planets, because the asteroids were presumed to have undergone the same planet formation process as the larger ones, and asteroids displayed evidences of the formation process that larger planets fail to display. Thus, taxonomy was functioning as a useful component of the scientific process, and there was no motive to reclassify asteroids.
6. By the 1950s, planet formation theories evolved to the point that they began to explain differences between asteroids and planets, so eventually the theoretical language bifurcated and theoreticians no longer called them *planets* except for the larger ones such as Ceres. The theoretical issue that made them distinct was at first the asteroids' presumed status as collisional fragments of protoplanets, then later it was the idea that they accreted as irregular bodies without ever having gone through the protoplanetary stage. Bodies that did go through a protoplanetary stage and were not subsequently disrupted, such as Ceres, were still called planets. The theoretical shift in terminology appears to have occurred precisely at the time of Kuiper's 1953 paper.
7. A flood of geophysical data became available in the latter half of the 20[th] century, evidenced by a great broadening of the topics in asteroid publications. This is empirically correlated to the timing in the late-1950s of an abrupt disappearance from the literature of papers that called asteroids *planets*. The timing of this sudden disappearance of usage correlates well to the paper by Kuiper in 1953 where theoreticians had also dropped the *planet* terminology, so this effectively marks when the community had reclassified asteroids as non-planets.

From this we conclude that the planetary science community did not reclassify asteroids on the basis of their sharing of orbits, which had been known in the extreme since the mid-19th century. Rather, they were reclassified beginning in the 1950s on the basis of new data showing asteroids' geophysical differences from large, gravitationally rounded planets, along with the theoretical developments that gave an adequate explanation for how asteroids came to possess such geophysical differences. In this process, there was never a point that the community argued Ceres is different than other planets, and the literature shows that scientists have continued to call it a planet to this day, citing geophysical reasons that are in-line with the well-attested historical developments. We therefore conclude that the argument made during the IAU planet definition controversy, that planet-sized Kuiper Belt Objects should be classified as non-planets because they share orbits, is arbitrary and not based on historical precedent.

We recommend that, regarding planetary taxonomy, central bodies such as the IAU do not resort to voting to create the illusion of scientific consensus. The IAU has done damage to the public perception of science, which is a process free from centrally dictated authority, in its imposition of a definition of *planet* and the number of planets fitting that definition, which has been instilled in educational textbooks around the world on the basis of their authority. Rather than voting on any other taxonomical issues, the IAU should simply rescind its planetary (and dwarf planetary) definitions and not replace them with any new definitions. In short, the IAU should simply allow the scientific communities to reach consensus on taxonomies through precedent set in literature and conference proceedings. We further recommend that educational organizations teach students that taxonomy is a vital part of science, along with observing nature, forming hypotheses, and testing predictions. Scientists utilize taxonomy to organize their observations of nature, to enable clearer thinking and to communicate concepts that they piece together into hypotheses. Therefore, definitions such as for *planet* are determined by that process, not arbitrarily nor to serve cultural purposes nor to fit comfortably with culture's preconceptions. The evolution of asteroid and planet taxonomy can be a pedagogical example of these concepts. We live in a time when the discovery of planetary bodies within our own solar system and around other stars is greatly expanding and revealing properties and solar system architectures not previously known or predicted. This will necessarily continue to drive the evolution of how we group objects into categories of planets and other taxa, motivated by scientific utility.